# Louth Crater: Evolution of a layered water ice mound

Adrian J. Brown[*,1,2], Shane Byrne[3], Livio L. Tornabene[3], and Ted Roush[2]

[1] *SETI Institute, 515 N. Whisman Rd, Mountain View, CA 94043, USA*
[2] *NASA Ames Research Center, Moffett Field, CA 94035, USA*
[3] *Lunar and Planetary Laboratory, University of Arizona, AZ, 85721, USA*

Corresponding author:
Adrian Brown
SETI Institute
515 N. Whisman Rd Mountain View, CA 94043
ph. 650 810 0223
fax. 650 968 5830
email. abrown@seti.org

Short running title: "Louth Crater: surface

## ABSTRACT

We report on observations made of the ~36km diameter crater, Louth, in the north polar region of Mars (at 70°N, 103.2°E). High-resolution imagery from the instruments on the Mars Reconnaissance Orbiter (MRO) spacecraft has been used to map a 15km diameter water ice deposit in the center of the crater. The water ice mound has surface features that include roughened ice textures and layering similar to that found in the North Polar Layered Deposits. Features we interpret as sastrugi and sand dunes show consistent wind patterns within Louth over recent time. CRISM spectra of the ice mound were modeled to derive quantitative estimates of water ice and contaminant abundance, and associated ice grain size information. These morphologic and spectral results are used to propose a stratigraphy for this deposit and adjoining sand dunes. Our results suggest the edge of the water ice mound is currently in retreat.

---

[*] corresponding author, email: abrown@seti.org

# INTRODUCTION

T he martian polar regions currently act as reservoirs for volatiles such as $CO_2$ and $H_2O$. Understanding the nature of these volatile ice deposits is critical to understanding the climate and its variability on modern Mars (Laskar et al., 2002). Ice deposits that are unusual in their properties or locations deserve extra scrutiny as they may reveal unexpected behavior that contributes to our understanding of the Martian climate. Here we report on observations of a 36km diameter crater in the north polar region, which bears the IAU provisional name of "Louth". Louth contains the southernmost large, exposed water ice deposit (at ~70°N). Korolev Crater and other perennial ice deposits in the northern polar region lie above 73°N. Here we interpret Mars Reconnaissance Orbiter (MRO) and Mars Odyssey images in order to address these questions:

- Can we determine the purity of exposed water ice, and can we model the water ice grain diameters - what does this tell us about the depositional history of the deposit?
- Are any changes taking place within the crater over short time scales (Martian weeks)?
- Why is the water ice mounds stable at this location?

**Imaging Instruments on Mars Reconnaissance Orbiter**

In this paper, we primarily use three of the imaging instruments onboard the MRO spacecraft, the High-resolution Imaging Science Experiment (HiRISE), the Compact Reconnaissance Imaging Spectrometer for Mars (CRISM) and Context Imager (CTX). The HiRISE is a high-resolution camera capable of providing red-filter (0.692μm) images at resolutions of 0.25-1.3m/pixel with a swath width of 6km (McEwen et al., 2007). HiRISE also has color filter coverage over an additional two wavelengths: blue-green (~0.536μm) and the near-IR (~0.874μm) over the center of the image with a swath width of ~1.2km. CRISM is a visible to near-infrared spectrometer sensitive to photons with wavelengths from ~0.4 to ~4.0μm (Murchie et al., 2007). It is well suited to observing water ice due to the strong absorption of $H_2O$ ice bands at 1.25, 1.5 and 2.0μm. In high-resolution mode, CRISM's instantaneous field of view (pixel) size corresponds to ~20m on the ground, with a swath width of ~12km. The CTX camera is a 5064 pixels-wide CCD attached to a Cassegrain (Maksutov) telescope, with pixel sizes of 6m on the surface of Mars. As the name implies, it obtains large images, 30km wide and up to 160km long, in one broad portion that includes much of the visible electromagnetic spectrum from 0.5-0.8μm (Malin et al., 2007).

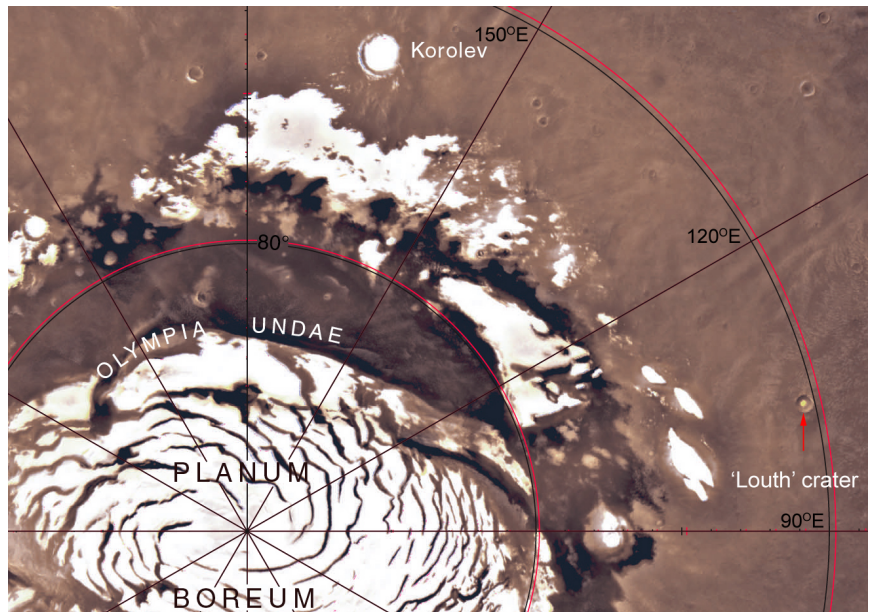

Figure 1. Viking MDIM map of the north polar region of Mars, showing the location of Louth Crater at 70°N, 103.2°E.

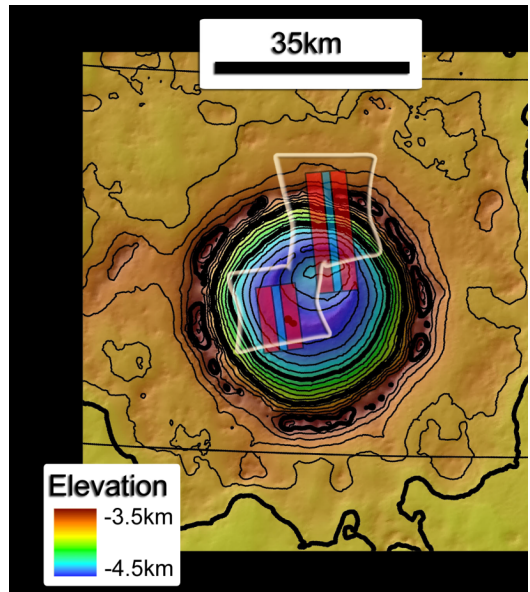

Figure 2. HiRISE and CRISM image footprints over a Mars Orbiting Laser Altimeter (MOLA) map of Louth Crater. The map is in polar stereographic projection, with the center of longitude = 103.2°E. All figures presented herein have the same projection.

## Louth Crater

Louth is a ~36km diameter, ~1.5km (rim to floor) deep (Figure 2) crater located at 70.5°N, 103.2°E which was observed in Viking images to contain a slightly elliptical, (~14km in the long dimension and ~12km in the short dimension) interior deposit of high albedo material (Figure 1). The water ice mound within Louth is the largest surface water ice deposit at this latitude. Its manner of deposition and apparent stability at this latitude are currently poorly understood.

This deposit was recently suggested to be water ice from summer temperature measurements taken by the Thermal Emission Imaging System (THEMIS) (Xie et al., 2006) showing mound temperatures of between 173-246K. In so doing, Xie et al. followed previous researchers who used thermal infrared temperature measurements to infer ice composition in the Martian poles (Kieffer et al., 1976; Byrne and Ingersoll, 2003; Titus et al., 2003). Xie et al. also reported a decrease in the size of the water ice mound between Martian fall and summer observations by the High-Resolution Stereo Camera (HRSC).

# OBSERVATIONS

Louth Crater was observed by MRO with two sets of coordinated observations during the first phase of science operations in the late northern summer on Mars. The coordinated observations are summarized in Table 1. CTX imagery (Figure 3) provides context for the HiRISE and CRISM images presented below.

| Instrument | File Name | $L_s$ | Orbit | Roll angle (°) | Incidence angle, i (°) | Phase angle, g (°) |
|---|---|---|---|---|---|---|
| HiRISE | PSP_001370_2505 | 133 | 1370 | 1.5 | 58.9 | 60.4 |
| | PSP_001700_2505 | 146 | 1700 | -8.6 | 64.2 | 56.0 |
| CTX | P01_001370_2503_ XI_70N257W_061111 | 133 | 1370 | 1.5 | 58.9 | 60.4 |
| | P02_001700_2506_ XI_70N256W_061206 | 146 | 1700 | -8.6 | 64.2 | 56.0 |
| CRISM | FRT00002F70_07 | 133 | 1370 | 1.53 | 58.8-59.3 | 48.8-83.5 |
| | FRT0000348E_07 | 146 | 1700 | -8.58 | 63.9-64.5 | 43.5-81.8 |

Table 1. High-resolution MRO images of Louth Crater used in this study. Sun incidence angle is relative to the standard Martian areoid.

## HiRISE observations of the water ice mound

Figure 4 shows six sub-images of the water ice mound taken from the two HiRISE observations of Louth Crater. These areas were selected to be illustrative of different surface types within the water ice mound.

*1. Rough 'stucco' water ice on the mound edge (Figure 4a).* This image shows the edge of the mound. The ice on the periphery of the mound displays a rough, disordered texture similar to a 'stucco' finish on adobe housing which is similar to the texture of the adjacent ice-free terrain. The area outside the edge of the main deposit shows meter-scale patches of high albedo material away from the continuous water ice mound. These patches decrease in number with increasing distance from the mound edge.

*2. Smooth bands with arcuate features (Figure 4b).* Arcuate features can be seen in the water ice mound at CTX scale. In

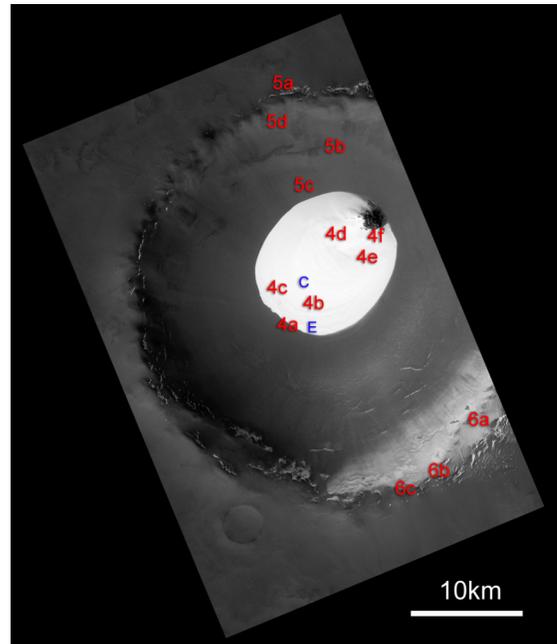

Figure 3. CTX image P01_001370_2503_XI_70N257W_061111 covering Louth Crater. Locations of images in Figure 4,5 and 6 are numbered in red. Locations where CRISM spectra were modeled are in blue – 'C' for mound center and 'E' for mound edge

HiRISE imagery these bands are seen to be smoother than the 'stucco' unit and contain linear features at the meter to decimeter scale (no slope corrections to derive the true layer thickness are available at this time).

*3. Exposure of stacked low albedo lines with dark material (Figure 4c).* In some locations, the arcuate features give way to a low albedo feature as seen in this figure. HiRISE imagery resolves an area of prominent stacked low albedo lines following the course of the arcuate feature, with dark albedo materials deposited along sinuous features that appear to emanate from the stacked low albedo lines at roughly equidistant intervals.

*4. Curvilinear and linear elongate mounds (Figure 4d).* Further into the interior of the water ice mound, an area of undulating topography is covered by high albedo, linear elongate mounds. In the CTX image (Figure 3 or 9) an undulating pattern is apparent with a period of roughly one kilometer. The large-scale undulation crestlines are orthogonal to the smaller linear to curvilinear elongate mounds in Figure 4d. These elongate mounds become less linear and more sinuous when in apparent troughs of the large-scale undulations, and adopt a dark streak to their top rim.

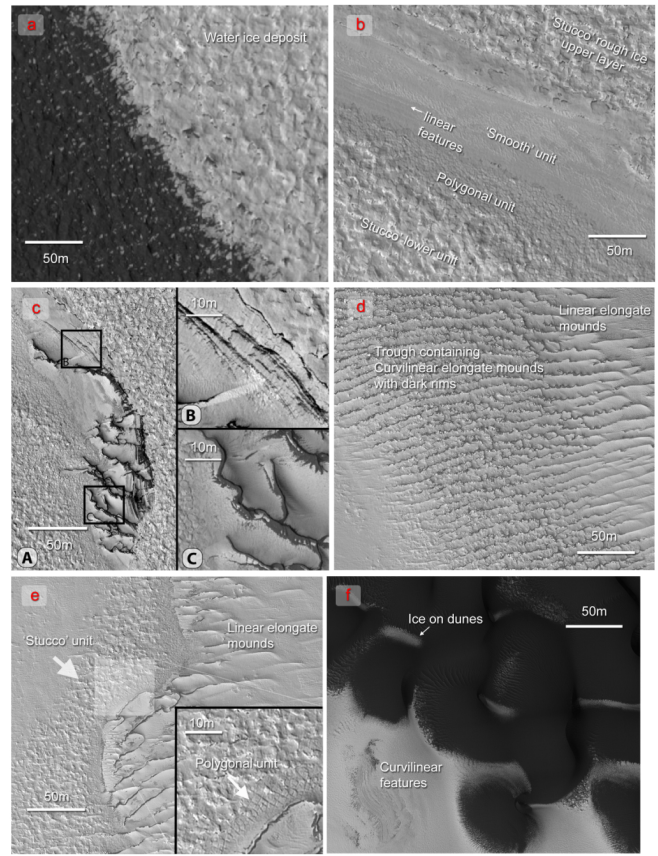

Figure 4. HiRISE red channel images of the water ice mound in Louth Crater. The first three are from HiRISE image PSP_001370_2505 and the last three from PSP_001700_2505. a.) Edge of the water ice mound b.) 'Smooth' layered unit. Note the smooth layers are not entirely smooth, but are smooth relative to the stucco water ice that dominates Figure 4a. A region of layering (or topographic steps in a homogenous material) is pointed out. c.) Smooth layered unit with dark material d.) 'Irregular' region showing trough filled with curvilinear and linear outside the trough e.) Polygonal layer along contact between sastrugi and smooth ice f.) Dunes and ice contact. Scale is identical to a.) in all images (unless marked), and projection is the same as Figure 3.

*5. Linear elongate mounds, polygonal unit and relatively smooth ice (Figure 4e).* This image shows relatively smooth ice on the top left side, followed by (from left to right) a rougher ice patch, then a polygonally patterned ice unit, followed by curvilinear dark rim elongate mounds, becoming more linear and less dark on the right side of the image. The sun is coming from the lower left in this image, and the inset captures what we interpret as a shadow cast (upwards in the inset) by the unit bearing the curvilinear dark rim elongate mounds.

*6. Contact between dunes and ice (Figure 4f).* The image shows the contact between the dark barchan dune deposit the edge of the ice mound. The dunes have been covered by small amounts of ice or frost in some locations. Some curvilinear features are also faintly visible in the water ice mound in this location.

**HiRISE observations of Louth Crater Basin, Rim and surroundings**

Figure 5 shows six sub-images from the two HiRISE observations of features seen in the crater basin and rim. These sub-images were chosen to highlight important morphologic units within the crater.

*1. High albedo deposits on the crater rim (Figure 5a)*. HiRISE observations show small patches of high albedo material on the outside crest of the rim of the crater. (Figure 5a). CRISM spectra of these locations show water ice absorption bands in pixels corresponding to the same locations where HiRISE resolves high albedo material. CRISM spectra of the low albedo soils surrounding these ice patches show no water ice bands at 1.25 and 1.5μm (also see Figure 7).

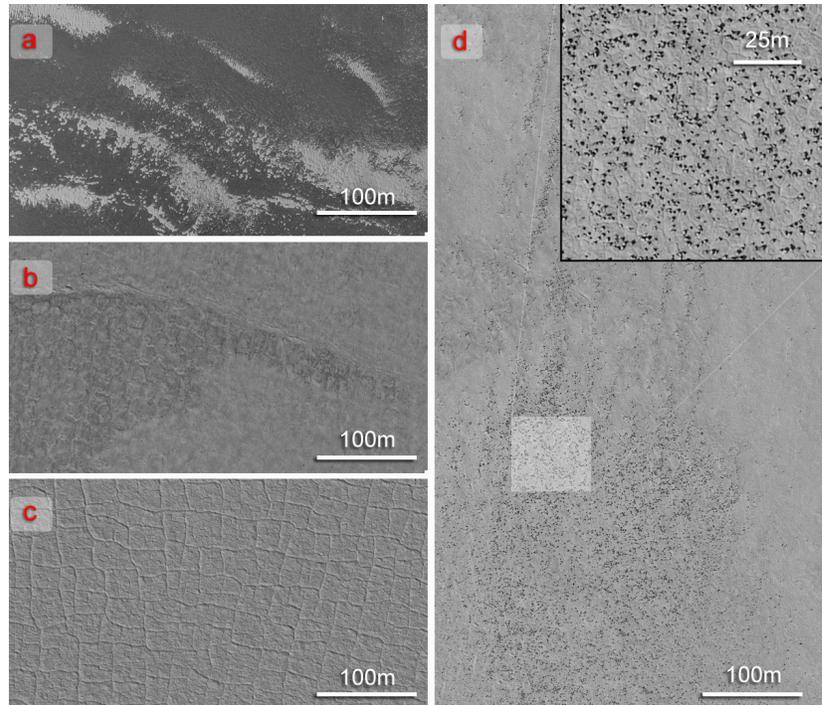

Figure 5. Four HiRISE sub-images from image PSP_001700_2505. These images show different features of Louth Crater away from the central ice mound. The locations of the sub-images are shown in Figure 3, the scale is the same in all images, and projection is the same as Figure 3. a.) Ragged high albedo patches confirmed as water ice rich by CRISM b.) High-resolution view of a low albedo circumferential feature within the crater basin. c.) Prominent quasi-rectangular polygonal ground in the crater basin close to the water ice mound d.) linear rows of boulders of similar dimension leading away from the crater rim (to the top of the image) to a circular boulder field. The inset shows the distribution of boulders and polygonal terrain beneath.

*2. Arcuate low albedo feature in crater basin (Figure 5b)*. An intriguing arcuate low albedo region shows the crater basin is not entirely homogenous. HiRISE resolves a small, discontinuous, raised rim on the outer (rim-ward) edge of this feature.

*3. Polygonal (quasi-rectangular) ground in crater basin (Figure 5c)*. Polygonal ground is present throughout the crater basin, and the scale of the polygons appears to be directly related to the proximity to the main water ice mound. Large-scale quasi-rectangular polygons are seen in this HiRISE image, which is relatively close to the water ice mound. The polygons decrease in size with increasing distance from the water ice mound.

*4. Boulder field on the crater wall (Figure 5d)*. The wall of the crater has a subdued appearance, and in one location 8 or 9 quasi-linear streams of meter-scale boulders extend from a 200m long section of the crater rim to an elliptical boulder field about 500m from the rim. The inset shows the meter scale boulders at higher resolution.

## CTX observations of albedo changes on the pole-facing crater rim

Two CTX observations were taken at Mars solar longitude ($L_s$) 133 and 146 to provide context for the HiRISE and CRISM imagery. We compared both CTX images for evidence of change over the intervening period. No changes to the central ice mound were found at the spatial resolution of CTX, however the poleward facing slope of the crater interior showed albedo changes in several locations over this time period (Figure 6). Most of the changes observed show parts of these areas getting darker as mid-summer ($L_s$=133) moves to later summer ($L_s$=146).

## CRISM observations of water ice

Figure 7 is a false-color image of the two high-resolution CRISM observations of Louth. The bands have been chosen to highlight the presence of water ice in light blue. In addition to water ice in the central mound, the CRISM data show water ice signatures on the crater rim. These crater rim water ice deposits were also imaged by HRSC on 02 Feb 2005 (Xie et al., 2006), indicating they are persistent features, or recurring seasonal features. These deposits correspond to the high albedo features in the HiRISE image shown in Figure 5a.

Slight differences in the color of the water ice mound between the two CRISM images are likely due to time dependent scattering by atmospheric aerosols or changes in the viewing geometry between the images. Atmospheric effects are not treated in this paper, with the justification that scattering from the atmosphere is less important than direct scattering from the surface in these high albedo regions. As a result, surface absorption bands in these ice-rich regions are much stronger than $CO_2$ gaseous absorption bands.

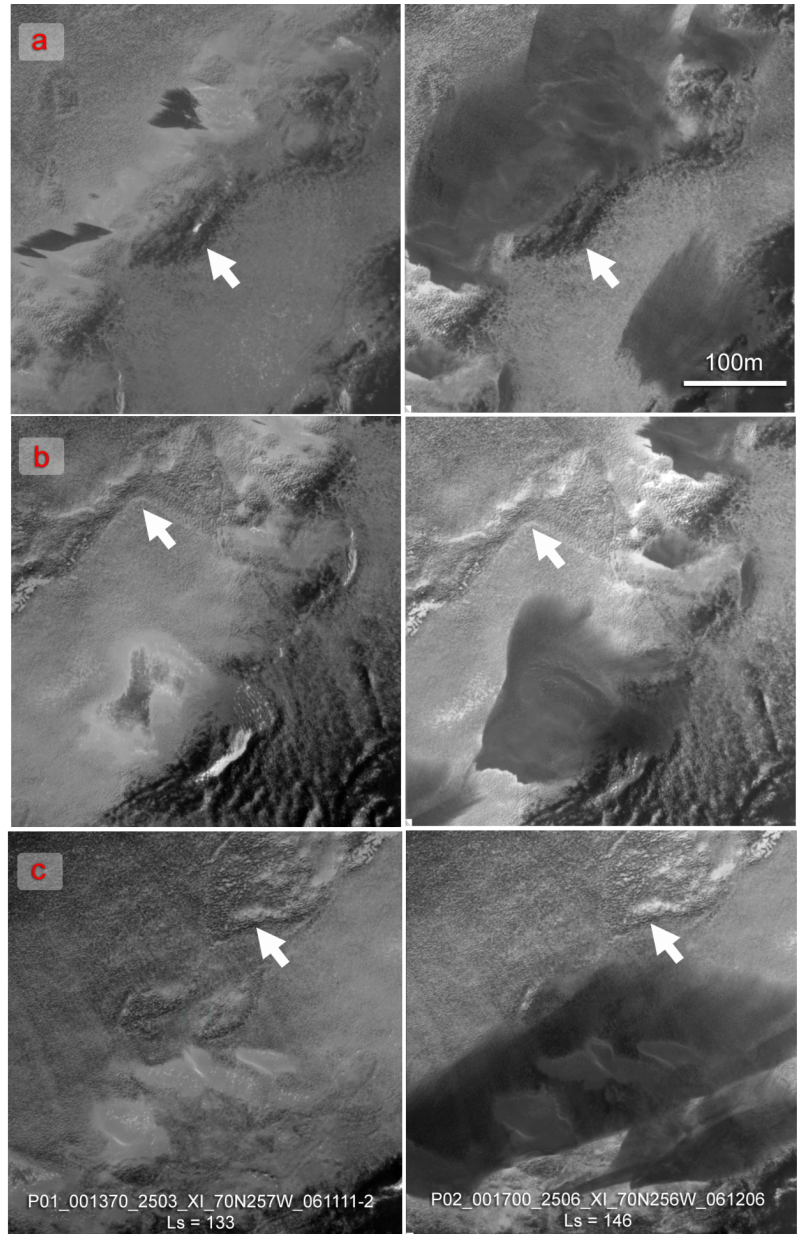

Figure 6. Three CTX sub-images of identical locations on the poleward facing crater rim of Louth Crater, showing changes in albedo between $L_s$ 133 (left) and $L_s$ 146 (right). Because the images are so different, we have used arrows to highlight features that appear similar at both periods. The locations of the sub-images are shown in Figure 3, the scale is the same in all images, and projection is the same as Figure 3. All of the images have been displayed with the same contrast stretching applied for reliable identification of differences.

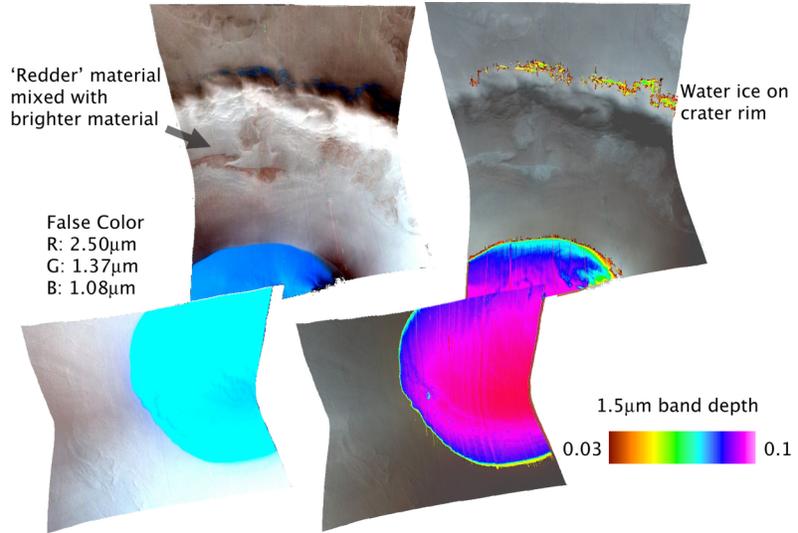

Figure 7. False-color mosaics of CRISM observations FRT00002F70_07 and FRT0000348E_07. The image on the left assigns Red Green and Blue channels to bands at 2.5, 1.37 and 1.08µm respectively. This shows water ice in light blue. The image on the right uses the inverse of this image as a base map and shows the band depth map of the 1.5 water ice band, showing how it varies over the scene. The band depth map was created using a band fitting technique (Brown, 2006) embedded in the software package 'MR PRISM' (Brown and Storrie-Lombardi, 2006).

The CRISM mosaic also shows subtle variations in albedo within the crater basin. Most of the basin is bright, but circumferential regions of lower albedo (red in this false-color representation) are present, corresponding to the low albedo region in Figure 5b.

**CRISM grain size modeling and contaminant modeling**

In order to get a quantitative assessment of the CRISM spectral information, we used a 'salt and pepper' single layer, bidirectional Hapke model to describe an intimate mixture of pure water ice and a palagonite (quenched basaltic tuff) contaminant component, as described in Roush (1994). The water ice optical constants at wavelengths less than 0.9µm were taken from Warren (1984) and greater than 0.9µm were taken from Grundy and Schmitt (1998). The palagonite data were taken from the combined Clark-Roush data profiled in Clancy et al. (1995). We describe further details of our modeling and error analysis in the Appendix.

We used equation 8.89 of Hapke (1993) to calculate the bidirectional reflectance. The equation is:

$$r(i,e,g) = \frac{w}{4\pi}\frac{\mu_0}{\mu_0+\mu}\left\{[1+B(g)]\,p(g)+H(\mu_0)H(\mu)-1\right\}$$

where r is reflectance, i the incidence angle, e is the emission angle, *g* is the phase angle, $\mu$ = cos(e), $\mu_0$ = cos(i), *p(g)* is the phase function, *w* is the single scattering albedo, *B(g)* is the backscattering function, and *H* is Chandresakhar's H-function for isotropic scatters (Chandresakhar, 1960). Because we lack data over a sufficient range of i, e and g, B(g) is given by:

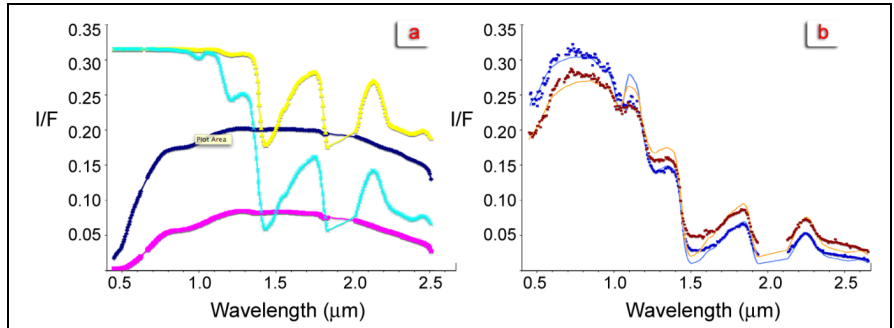

Figure 8. a.) Synthetic spectra of pure components modeled using optical constants – (yellow) $H_2O$ ice of 10 micron grain size (cyan) $H_2O$ ice of 100 micron grain size (dark blue) palagonite of 10 micron grain size (purple) palagonite of 100 micron grain size.
b.) CRISM spectra of water ice at the center (blue) and edge (brown) of the central mound in Louth Crater (dots) with modeled spectra of intimate mixtures of water ice and contaminant mixtures (lines). Discrepancies between models and data at ~1.1µm may be due to $H_2O$ optical constant discrepancies – see discussion in Grundy and Schmitt (1998)

$$B(g) = 1/(w(1+(\tan(g/2))/h))$$

where $h$, related to the opposition surge width (Hapke, 1986) is set to 0.05 (a lunar-like surface). A rigorous determination of $h$ requires observations at several phase angles and this is beyond the scope of this study. We also assume isotropic scattering, i.e. $p(g)=1$.

The optical constants are used to calculate $w$ as described in Roush (Roush, 1994). Our modeling employed two compositional materials, water ice and palagonite, and we considered several mixtures of these. The grain sizes and relative mass fractions were free parameters. A non-linear least squares fit, between the observed and calculated spectra, was minimized to find the best spectral match.

Figure 8 presents CRISM spectra from the edge and interior (see Figure 3 for locations) of the water ice mound (dots) and the best-fitting models (lines). The edge sample was taken from a rough 'stucco ice' region similar to that seen in Figure 4a. The central sample was taken from a smooth region similar to that seen in the top left corner of Figure 4e.

| Model | $n$ | component(s) | relative mass fraction (%) | grain diameter (mm) | $\chi^2$ |
|---|---|---|---|---|---|
| 1a | 4 | ice/ice/ice/ice | 58/32/10/<1 | 0.85/0.80/0.17/0.009 | 0.1101 |
| 1b | 3 | ice/ice/ice | 35/64/~1 | 1.92/0.78/0.012 | 0.1025 |
| 1c | 2 | ice/ice | 81/19 | 1.68/0.22 | 0.1030 |
| 1d | 1 | ice | 100 | 0.58 | 0.1207 |
| **2** | **4** | **ice/ice/plg/plg** | **43/57/<1/<1** | **1.97/0.49/0.46/0.002** | **0.0187** |
| 3 | 3 | ice/ice/plg | 4/94/2 | 1.05/0.48/0.46 | 0.0434 |
| 4 | 2 | ice/plg | 98/2 | 0.49/0.50 | 0.0443 |
| 5 | 1 | plg | 100 | 0.031 | 4.4136 |
| 6 | 2 | plg/plg | 100/<1 | 1.95/*0.00009* | 3.3123 |
| 7 | 3 | plg/plg/plg | 82/18/<1 | 1.18/0.38/*0.000008* | 3.3810 |

Table 2. Compositional models for Ice Mound Center. $n$ is the number of components. 'plg' indicates palagonite. All models have a fitting tolerance of $10^{-3}$. *Italics* indicate grain diameters violating the geometric optics assumptions of Hapke theory. **Bold** indicates the fit converging to the smallest $\chi^2$ value.

Tables 2 and 3 summarize the various best-fit models. We found models with pure palagonite (up to 3 grain sizes) or water ice (up to four grain sizes) were poor fits to the spectra, indicating neither pure water ice nor pure palagonite dominates. The best fits were achieved with models that contained two water ice and two palagonite components.

| Model | *n* | component(s) | relative mass fraction (%) | grain diameter (mm) | $\chi^2$ |
|-------|-----|--------------|---------------------------|---------------------|----------|
| 1a | 4 | ice/ice/ice/ice | 48/12/39/<1 | 1.14/0.90/0.16/0.0017 | 0.4563 |
| 1b | 3 | ice/ice/ice | 74/24/2 | 2.03/1.381/0.023 | 0.3552 |
| 1c | 2 | ice/ice | 85/15 | 70.4/0.19 | 0.3367 |
| 1d | 1 | ice | 100 | 0.44 | 0.4427 |
| **2** | **4** | **ice/ice/plg/plg** | **84/13/~2/<1** | **0.34/0.46/0.39/0.033** | **0.0395** |
| 3 | 3 | ice/ice/plg | 57/39/4 | 1.58/0.20/0.43 | 0.0539 |
| 4 | 2 | ice/plg | 98/2 | 0.33/0.13 | 0.0454 |
| 5 | 1 | plg | 100 | 0.031 | 3.0132 |
| 6 | 2 | plg/plg | 100/<1 | 0.84/*0.00007* | 2.2289 |
| 7 | 3 | plg/plg/plg | 6/94/<1 | 2.12/0.74/*0.00006* | 2.2364 |

Table 3. Compositional models for Ice Mound Edge. *n* is the number of components. 'plg' indicates palagonite. All models have a fitting tolerance of $10^{-3}$. *Italics* indicate grain diameters violating the geometric optics assumptions of Hapke theory. **Bold** indicates the fit converging to the smallest $\chi^2$ value.

Our modeling suggests higher amounts of contaminant (palagonite) and smaller water ice grain sizes are present near the edge of the ice mound when compared to the center of the mound.

# DISCUSSION

## Interpretation of Grain Size Results

Kieffer (1990) estimated the rate of ice grain growth under Martian conditions through thermal metamorphism. Assuming that all ice grains are deposited with similar dimensions, that their growth is not impeded or assisted by impurities and that all parts of this deposit experience a comparable thermal history, our grain size results suggest that ice exposed at the center of this deposit is older than that exposed in the edge. Grain size information can be used to infer stratigraphic relationships between the water ice units.

## Evolution of the Louth water ice mound

Using grain size, surface texture and contacts between units, we have constructed an interpretative map of the water ice mound, shown in Figure 9. The units of this map are discussed below in order from oldest to youngest.

Unit 1. *Smooth undulating layer in the center of the mound*. Assuming the water ice mound has a uniform temperature history, the relatively large grain size suggests the central part of the mound is the oldest. It is likely that the larger surface grains observed in this unit have been exhumed as a result of recent ice sublimation or aeolian activity.

We interpret the decimeter-scale linear to curvilinear elongate mounds to be wind blown ice mounds (or sastrugi). Sastrugi are sharp, irregular grooves formed on a snow surface by wind erosion and deposition (Grey and Male, 2004). We do not have a good understanding of the variations in albedo of the sastrugi and the nature of the low albedo rims. The 'dark top rims' are composition related and not caused by shadows because they cover both sides of the tops of elongate mound rims. Whether the sastrugi have interiors that predominantly consist of high albedo ice or low albedo material is an open question that this stage.

Sastrugi differ from sand dunes because their ridges are parallel to wind direction (Grey and Male, 2004). The long dimension of the sastrugi mounds points in the direction of the barchan dunes of unit 2 (discussed below), and orthogonal to the kilometer scale undulations. We consider it likely the prevailing winds have developed all these features.

Unit 2. *Dark sand dunes*. Water ice has been deposited after the dark dune deposit was formed. It is not clear whether water ice was present before (or during) dune formation. Multiple possible histories for formation for the sand dunes exist:

A.) the dunes formed before the ice mound, which formed around them,

B.) the dunes formed from material that was previously trapped in layers of a larger water ice mound that has since sublimed.

C.) the dunes formed after the ice mound, and small amounts of ice in and on the dunes are just a thin frost.

Unit 3. *'Irregular' region and linear to curvilinear elongate mounds*. An area of varying topography, containing light linear and dark rimmed curvilinear features we also interpret as sastrugi, lies closest to the dune deposit and may be influenced by the same aeolian wind patterns that formed the exposed dark sand dunes. Linear sastrugi are oriented perpendicular to the dune slipfaces, indicating a consistency in wind patterns across Louth that has transcended dune and water ice mound formation.

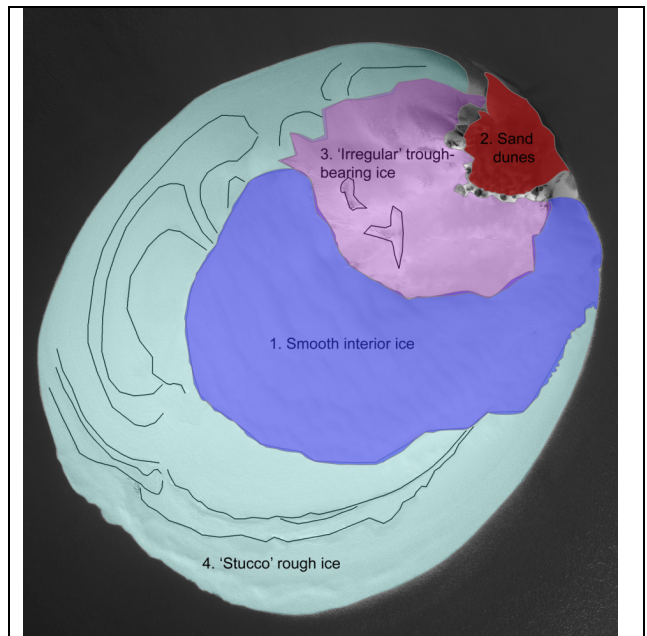

Figure 9. An interpretive map of the water ice mound at the center of Louth Crater, showing (in order from oldest to youngest) 1.) smooth, interior ice with kilometer-scale undulations (blue) 2.) low albedo sand deposits (red) 3.) 'Irregular' trough and satrugi-bearing ice (purple) 4.) 'stucco' rough water ice deposits at the edge (green). Smooth, layered arcuate features have been outlined in dark green.

Unit 4. *'Stucco' ice edge deposits.* Based on relatively small grain size, we believe the 'stucco' edge deposits represent the youngest regions of the water ice mound. The similarity in texture to the adjacent ice-free terrain (Figure 4a) indicates that the ice cover is thin.

The arcuate features in the relatively smooth unit highlighted in Figure 4b (and outlined in Figure 9) appear to coincide with breaks in slope and perhaps indicate steps or layers that track the period growth patterns of the water ice mound, in a manner analogous to the layers of the North Polar Layered Deposits (NPLD) (Milkovich et al., 2007). Future high-resolution topographic information will be needed to confirm this interpretation.

In a previous survey of northern polar craters, Garvin et al. used MOLA data to measure height profiles of craters (Garvin et al., 2000). Although they did not specifically mention Louth crater, they observed the presence of layered deposits in craters at 82N,190E and 77N,89E (their 'A' and 'B' features) and also likened them to the NPLD. Garvin et al. postulated that crater infill deposits (including ice covered mounds such as those at Louth) would take millions of years to develop at the polar sedimentation rates suggested by Thomas et al. (1992) and therefore they may have been created by episodic increased sedimentation associated with the advance of the polar ice cap and later preserved by the relative protection of the crater rim from further erosion.

Scattered icy outliers around the edge of the water ice mound (Figure 4a) may be remnants of a previously larger extent of the ice deposit, supporting the notion that it is currently losing mass.

**Crater rim and interior materials**

The HiRISE image in Figure 5d shows well-sorted boulder fields in quasi-linear arrangements down the slope of the crater rim. A 'pool' of such boulders has formed inside the crater basin. The linear arrangement and sorted nature of these boulders might be explained by mass wasting that carried them down to a low point within the basin. The linear arrangements of boulders may also indicate preferred downhill paths the boulders have followed that are difficult to distinguish in the HiRISE image.

The low albedo circumferential region in the crater basin seen in Figure 5b appears to be bounded by a continuous top edge that is also associated with a discontinuous low rise. The origin of this low albedo region is unclear – one possibility is that it was formed by stripping back of brighter material to reveal hummocky ground beneath – another possibility is that brighter material surrounding it masks a much larger extent of the low albedo circumferential material. CRISM reveals no spectral difference (other than the albedo change) between the low albedo and adjoining bright materials in this region.

Large prominent polygons are seen in the interior of the crater adjacent to the ice mound (Figure 5c), with decreasing size away from the center of the crater. Polygon size is related to release of tensile strength in the ice-cemented regolith, with smaller polygons resulting from higher stresses (Mellon, 1997). The polygon size variations we see on the crater floor could therefore be related to variation in the stress field produced when this terrain cools during the winter. As albedo and latitude are not varying between these features the cooling rate is governed by the thermal inertia of the upper meter. Smaller polygons may result from lower thermal inertia that could possibly be due to lower regolith ice content or a deeper ice table in these locations.

**Albedo changes on the pole-facing crater rim**

CTX imagery shows the pole-facing slope of Louth crater displays changes in albedo between $L_s$ 133 and 146. Three plausible scenarios may explain these changes:

    A.) Ice sublimation (and, to a lesser extent, deposition)
    B.) Ice grain growth
    C.) Aeolian controlled impurities (deposition of dust)

Multispectral, medium spatial resolution (~100m/pixel) CRISM data obtained at $L_s$ 160 (image name MSP00003B9C_05) shows water ice as the dominant spectral component on the poleward facing crater interior at that time (K. Seelos, personal communication 2007). $CO_2$ ice is not thermodynamically stable at the temperatures shown in Figure 10 and is unlikely to have contributed to these surface changes in albedo.

We consider $H_2O$ ice grain growth unlikely in such a short period, but scenarios A.) and C.) remain plausible explanations. We cannot discriminate between these two scenarios without several CRISM full resolution images over this region during the northern Martian spring and summer. We intend to investigate this rim with full resolution CRISM data as the MRO mission continues.

**Formation and stability and future of the Louth water ice deposit**

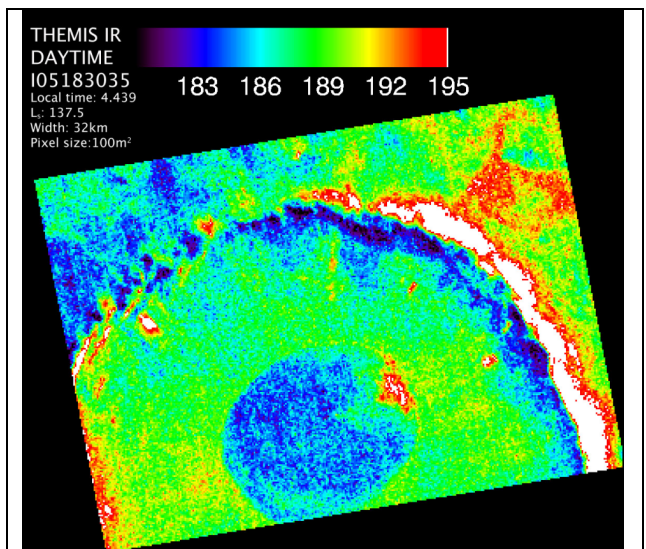

Figure 10. THEMIS IR Daytime view of Louth Crater at $L_s$ 137.5. Colors correspond to brightness temperatures at local time of 4.25am, at time of Mars Odyssey overflight. Note 'warm' patch corresponding to dark dunes, with some relatively warm border areas extending into the 'irregular' unit (purple in Figure 9). When this image was collected, the sun was approximately 8 degrees above the horizon - this is what is causing the elevated temperatures in the upper right corner.

<u>Formation.</u> Factors influencing the initial formation of the water ice mound in Louth crater remain a mystery. Its location and relative proximity to the radial water ice deposits that lie around the north polar cap between longitudes 90-250° and latitudes 75-78° (unofficially named "Mrs. Chippy's Ring" (Calvin and Titus, 2004)) suggests linkages to these off-cap water ice deposits. It is unlikely that direct transfer of icy material by aeolian activity from "Mrs. Chippy's Ring" to Louth crater (i.e. saltation of ice grains) is possible given the intervening 100's of kms. There is little evidence to suggest Louth is a remaining outlier from a much larger polar cap. Therefore our preferred mode of deposition is via atmospheric deposition of water vapor, where water particles are preferentially trapped in the center of the crater. In competition with this depositional process, aeolian activity and sublimation also appears to be shaping and stripping back the water ice mound, and it is possible, given the offset position of the water ice mound, that it formerly occupied a larger area within Louth and has retreated poleward in recent times.

<u>Stability.</u> Taking the ice mound to be a section of a sphere that is 12km across and 200m deep, the water ice mound deposit is roughly 11km$^3$ in volume. Its location in the crater is offset to the poleward side. Vasavada et al. used thermal modeling to suggest craters on the Moon and Mercury at similar latitudes will experience coldest average temperatures on the crater floor adjacent to the equatorward rim (Vasavada et al., 1999). However the depth to diameter ratio of Louth crater is only ~0.42 so the walls appear to be less than 5 degrees in height when standing at the craters deepest point. Mars also has a much larger obliquity than either the Moon or Mercury so that the sun may appear up to 40 degrees above the horizon at this location. These two properties mean direct shadowing by crater walls of the crater floor is not effective at Louth. The higher albedo and thermal inertia of the ice helps stabilize it against ablation, but this does not explain why the icy material began forming in this location. We believe we can therefore narrow the factors controlling on the size and shape of the water ice mound to:

A.) The shape of Louth crater exerts control over local meteorology and therefore the location of the water ice mound, or

B.) Thermodynamic properties of the material underlying the ice deposit determines its location.

<u>Future.</u> The current state of the icy deposit is likely to be one of retreat. The small outliers in Figure 4a could be interpreted as thermodynamically stable patches caused by poleward facing slopes within the irregular topography of the crater floor. However, the fact that the number of these outliers decreases as one moves away from the edge of the main icy deposit (despite the fact that the texture of the crater floor is unchanged) suggests that these patches are instead remnants of a once larger extent of the main mound.

Results of the spectral modeling suggests that the water ice is 99% pure at the optical surface – however, it is possible there is a greater proportion of dark material trapped within the mound, perhaps in the form of interior layers linked to those that are visible around the edges.

# ACKNOWLEDGEMENTS

We would like to thank the entire MRO Team, particularly the PI's for HiRISE (Alfred McEwen), CRISM (Scott Murchie) and CTX (Mike Malin) and their staff. Special thanks to Kim Seelos for her help in identifying water ice in CRISM multispectral images. We would like to thank two anonymous reviewers for their extensive suggestions that improved the paper immensely. Support provided by the MRO program enabled the participation of TLR.

# APPENDICES

## Appendix 1 - CRISM VNIR and IR spectrum splicing and calibration state

As described in the text, the CRISM spectra are measured by separate VNIR (0.4-1.0μm) and IR (1.0-4.0μm) detectors. In order to get a full spectrum from 0.4-4.0μm, the modeled spectra in Figure 8 were spliced together by hand - the overlapping channels that best matched in I/F were selected and other overlapping points were discarded. Spectral points between 1.9-2.1μm that were affected by a strong $CO_2$ gas triplet absorption band at 2.0μm were removed. VNIR version 2 (TRR2) and IR version 0 calibrated data were used. CRISM experiences a change in detector responsiveness across the imaging slit, which means that central wavelengths of pixels in the center of the image are lower than pixels in the same band on the edge of the image (this is known as a 'spectral smile'). This effect was removed by using a look up table that plotted the central wavelength of each detector column – this lookup table is termed a 'CDR_4' file by the CRISM team and we used CDR410803692813_WA_0000000S_2.IMG and CDR410803692813_WA_0000000L_2.IMG for the VNIR and IR detector, respectively.

## Appendix 2 - Hapke modeling error analysis

We investigate the uncertainty in derived model parameters by starting the fitting procedure with several different initial estimates for the abundances and grain sizes of the various components. We restrict our sensitivity analysis to those models where the chi-squared values are within ~10% of the minimum chi-squared value.

Our best fitting spectral mixture is comprised of two different ice and palagonite members, each with different grain sizes. For the ice mound edge ('E' in Figure 3) models we find that at the 3 sigma level the ice 1 relative mass fraction (RMF) varies by <0.3% of the mean value (0.84) and diameter by ~9% of the mean value (0.33 mm). For ice 2, at the 3 sigma level, the RMF varies by 0.8% of the mean value (0.13) and diameter by ~2% of the mean value (0.46 mm). For Palagonite 1, at the 3 sigma level, the RMF varies by ~3% of the mean value (0.025) and diameter by ~3% of the mean value (0.39 mm). For Palagonite 2, at the 3 sigma level, the RMF varies by 33% of the mean value (0.005) and diameter by <0.2% of the mean value (0.03 mm).

For the Ice Mound Center ('C' in Figure 3) models we find that at the 3 sigma level the ice 1 RMF varies by ~4% of the mean value (0.43) and diameter by ~5% of the mean value (2.00 mm). For ice 2, at the 3 sigma level, the RMF varies by ~3% of the mean value (0.57) and diameter by ~4% of the mean value (0.50 mm). For Palagonite 1, at the 3 sigma level, the RMF varies by 60% of the mean value (0.037) and diameter by ~4% of the mean value (0.47 mm). For Palagonite 2, at the 3 sigma level, the RMF varies by 16% of the mean value (0.005) and diameter by ~20% of the mean value (0.0017 mm).